\input{aipcheck.tec}

\documentclass[
    ,final            
  ]
  {aipproc}
 
\layoutstyle{6x9}

\begin{document}

\title{All Possible Lightest Supersymmetric Particles in R-Parity Violating mSUGRA Models and their Signals at the LHC}


\author{H.~K.~Dreiner}{
  address={Physics Institute, University of Bonn, Bonn, Germany}
}

\author{S.~Grab}{
  address={Santa Cruz Institute for Particle Physics, University of California, Santa Cruz, CA 95064, USA}
}

\begin{abstract}
We consider minimal supergravity (mSUGRA) models with an additional R-parity violating operator 
at the grand unification scale. This can change the supersymmetric spectrum leading on the one 
hand to a sneutrino, smuon or squark as the lightest supersymmetric particle (LSP). 
On the other hand, a wide parameter region is reopened, where the scalar tau is the LSP. 
It is vital to know the nature of the LSP, because supersymmetric particles normally cascade 
decay down to the LSP at collider experiments. We investigate in detail the conditions leading 
to non-neutralino LSP scenarios. 
We also present some typical LHC signatures.
\end{abstract}

\maketitle


\section{Introduction}

In the minimal supersymmetric standard model (MSSM) the LSP 
is stable guaranteed by the discrete symmetry R-parity, $R_p$. The LSP
must then be the lightest neutralino for cosmological reasons.
However, if we drop $R_p$, further renormalizable operators are allowed in the
superpotential \cite{Allanach:2003eb}
\begin{equation}
W_{\not R_p}= \frac{1}{2} \lambda_{ijk} L_i L_j \bar E_k +
\lambda'_{ijk} L_i Q_j \bar D_k + \frac{1}{2} \lambda''_{ijk} \bar U_i 
\bar U_j \bar D_k + \kappa_i L_i H_d \, . 
\label{superpot}
\end{equation}
To ensure the stability of the proton we must either suppress the lepton-
or the baryon-number violating operators in Eq.~(\ref{superpot}).
These terms violate $R_p$ and thus the LSP is no longer stable
\footnote{Potential other dark matter candidates are for example
the axino or the lightest U-parity particle \cite{Chun:1999cq}.}. 
It can be in principle any supersymmetric particle (sparticle) \cite{Dreiner:2008ca} 
\begin{equation}
\tilde\chi^0_1,\;\tilde\chi^\pm_1,\;\tilde\ell^\pm_{L/Ri},\;\tilde\tau_1,\;
\tilde\nu_i,\;\tilde q_{L/Rj},\;{\tilde b_1}, \;\tilde t_1,\;\tilde g\,.
\label{LSPs}
\end{equation}
It is vital to know the nature of the LSP, because sparticles normally 
cascade decay down to the LSP at collider experiments. 

However, due the bewildering array of LSP candidates in Eq.~(\ref{LSPs}), it is difficult
to perform detailed phenomenological studies. We therefore need a guiding principle. 
As a first step, we investigate the $R_p$ violating (${\not R_p}$) minimal supergravity 
model \cite{Allanach:2003eb}:
\begin{equation}
M_0 \, , M_{1/2} \, , A_0 \, , \tan \beta \, , {\rm sgn}(\mu) \, ,
\Lambda \,  \quad \quad {\rm with} \quad 
\Lambda \in \{\lambda_{ijk},\lambda'_{ijk},\lambda''_{ijk} \} \, . 
\label{mSUGRA_param}
\end{equation}
Here, $M_0$ ($M_{1/2}$) is the universal softbreaking scalar
(gaugino) mass and $A_0$ is the universal softbreaking
trilinear scalar interaction at the grand unification scale
$M_{\rm GUT}$. $\tan \beta$ is the ratio of the 
two vacuum expectation values and $\mu$ is the Higgs 
mixing parameter. The effects of ${\not R_p}$ 
are incorporated by assuming one non-vanishing 
trilinear coupling $\Lambda$ at $M_{\rm GUT}$ at a time,
{\it cf.} Eq.~(\ref{superpot}).

We now have a simple and well-motivated framework, in which we
can systematically investigate the nature of the LSP and its
phenomenology at the LHC. 

\section{LSP Candidates in $R_p$ violating mSUGRA models}

\begin{figure}
\begin{tabular}{cc}
 \centering
    {
        \includegraphics[width=7.3cm]{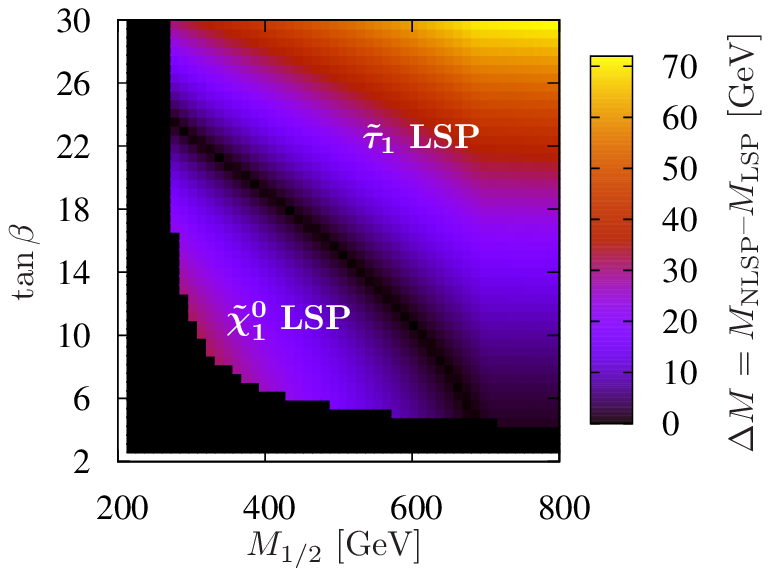}
		\put(-15.0,15.0){\makebox(0,0){(a)}}
    }
    &
    {
        \includegraphics[width=7.3cm]{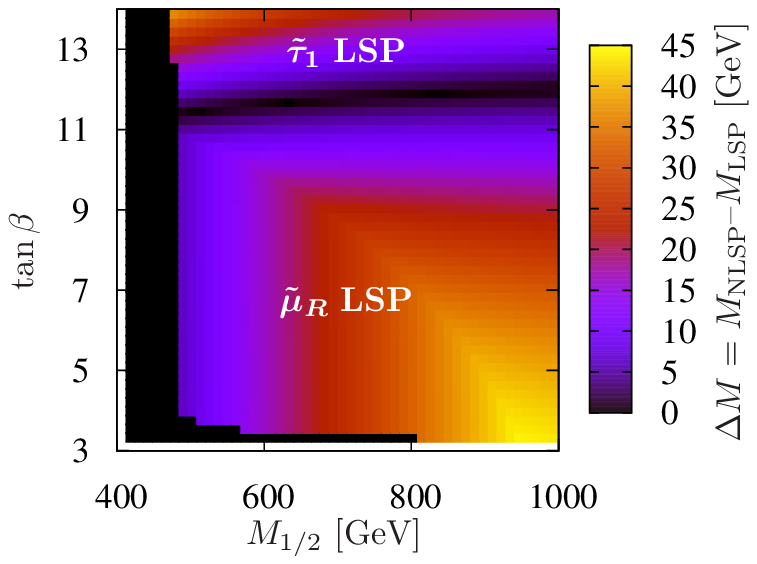}
        \put(-15.0,15.0){\makebox(0,0){(b)}}
    }
    \\
    {
        \includegraphics[width=7.3cm]{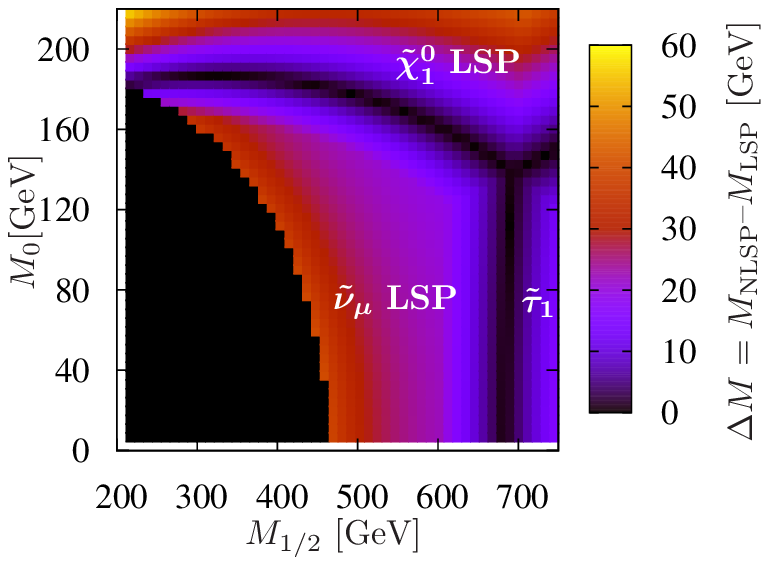}
        \put(-15.0,15.0){\makebox(0,0){(c)}}
    }
    &
    {
        \includegraphics[width=7.3cm]{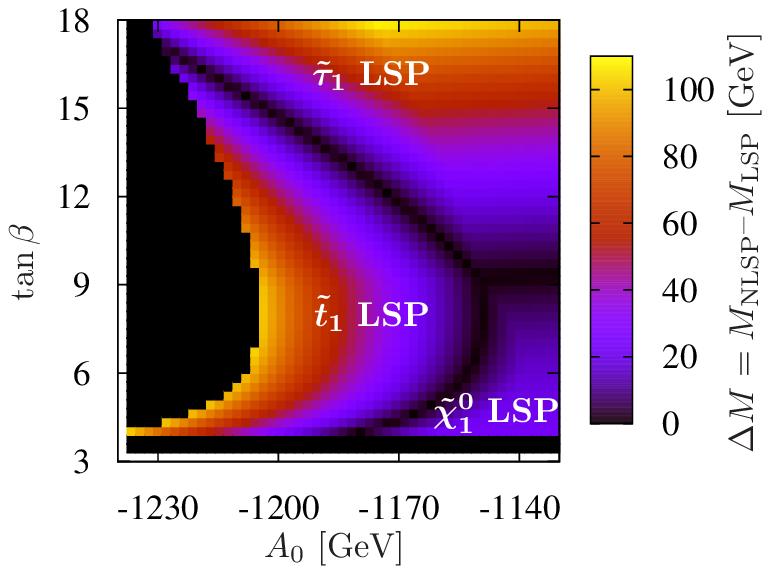}
        \put(-15.0,15.0){\makebox(0,0){(d)}}
    }
\caption{\small Mass difference, $\Delta M$, between the next-to-lightest LSP 
(NLSP) and LSP. The LSP candidates are explicitly mentioned. The blackened out region 
corresponds to parameter points, which posses a tachyon or which violate other
experimental constraints; 
see Ref.~\cite{Dreiner:2008ca,Bernhardt:2008jz}
for details. {\bf Fig.\ref{LSP_candidates}(a)}: $\tilde{\tau}_1$ LSP region;
$M_0=100$ GeV, $A_0=-100$ GeV, ${\rm sgn}(\mu)=+1$, $\Lambda < \mathcal{O}(10^{-1})$. 
{\bf Fig.\ref{LSP_candidates}(b)}: $\tilde{\mu}_R$ LSP via $\lambda_{132}|_{\rm GUT} = 0.09$;
$M_0=170$ GeV, $A_0=-1500$ GeV, ${\rm sgn}(\mu)=+1$. 
{\bf Fig.\ref{LSP_candidates}(c)}: $\tilde{\nu}_\mu$ LSP via $\lambda'_{231}|_{\rm GUT}=0.11$;
$A_0=-600$ GeV, $\tan \beta=$ 10, ${\rm sgn}(\mu)=+1$. 
{\bf Fig.\ref{LSP_candidates}(d)}: $\tilde{t}_1$ LSP via $\lambda''_{323}|_{\rm GUT}=0.35$;
$M_0=120$ GeV, $M_{1/2}=480$ GeV, ${\rm sgn}(\mu)=+1$.
 \label{LSP_candidates}}
\end{tabular}
\end{figure}

In most of the ${\not R_p}$ mSUGRA parameter space the
lightest neutralino $\tilde{\chi}_1^0$ or the 
lightest scalar tau $\tilde{\tau}_1$ is the LSP
as can be seen in Fig.~\ref{LSP_candidates}a
\footnote{All sparticle mass spectra have been calculated
with {\tt SOFTSUSY3.0} \cite{softsusy}.}. 

According to Fig.~\ref{LSP_candidates}a, we can obtain a $\tilde{\tau}_1$ 
LSP instead of the $\tilde{\chi}_1^0$ LSP by increasing
$M_{1/2}$. Increasing $M_{1/2}$ increases the mass of the 
(bino-like) $\tilde{\chi}_1^0$ faster than the mass of the 
(mainly right-handed) $\tilde{\tau}_1$. Apart from that, we 
can also get a $\tilde{\tau}_1$ LSP by increasing $\tan \beta$. 
Increasing $\tan \beta$ increases on the one hand the magnitude 
of the tau Yukawa coupling, which increases its (negative) effect
on the running of the $\tilde{\tau}_1$ mass~$^3$. On the other hand, $\tan \beta$ increases
the mixing between $\tilde{\tau}_L$ and $\tilde{\tau}_R$. We conclude
that a $\tilde{\tau}_1$ LSP is as well motivated as a $\tilde{\chi}_1^0$
LSP in ${\not R_p}$ mSUGRA.

We have assumed in Fig.~\ref{LSP_candidates}a that 
$\Lambda$ at $M_{\rm GUT}$ is at least one order of magnitude smaller than
the gauge couplings, {\it i.e.} it has no significant impact
on the running of the sparticle masses. However,
for $\Lambda = \mathcal{O}(10^{-1})$, we need to take into account
the ${\not R_p}$ contributions to the renormalization group equations (RGEs).
If a sparticle directly couples to $\Lambda$, the dominant contributions
to the RGE of the running sparticle mass $\tilde{m}$ are 
\cite{Allanach:2003eb,Dreiner:2008ca,Bernhardt:2008jz}
\footnote{For third generation sparticles we also need to take 
into account the contributions from the Higgs-Yukawa interactions. 
Their effect is similar to $\Lambda$ and $h_\Lambda$ in Eq.~(\ref{RGE}).}:
\begin{equation}
16\pi^2 \frac{d (\tilde{m}^2)}{dt} = 
- a_i g_i^2 M_i^2 - b g_1^2 \mathcal{S} + \Lambda^2 \mathcal{F}
+ c\, h^2_\Lambda \;, \quad
h_\Lambda \equiv \Lambda \times A_0  \,\,\, {\rm at} \,\,\, M_{\rm GUT}\,.
\label{RGE}
\end{equation}
Here $g_i$ ($M_i$), $i=1,2,3$, are the gauge couplings (soft
breaking gaugino masses). $t=\ln Q$ with $Q$ the renormalization
scale and $a_i$, $b$, $c$ are constants of
$\mathcal{O}(10^{-1}-10^1)$. $\mathcal{S}$ and $\mathcal{F}$ are linear
functions of products of two softbreaking scalar masses.

The sum of the first two ${\not R_p}$ terms in Eq.~(\ref{RGE}) is negative
and thus \textit{increases} $\tilde{m}$ when running from $M_{\rm
GUT}$ to the electroweak scale. In contrast, the last two ${\not R_p}$ terms
proportional to $\Lambda^2,\,h^2_\Lambda$, are
always positive and therefore \textit{decrease} $\tilde{m}$. We thus
expect new LSP candidates beyond $\tilde{\chi}^0_1,\,\tilde\tau_1$ if these
latter terms contribute substantially, {\it i.e.} $\Lambda=\mathcal{O}(g_i)$ 
\cite{Dreiner:2008ca,Bernhardt:2008jz}. We can strengthen the (negative) 
contribution of $h^2_\Lambda$, by choosing a negative $A_0$ with a 
large magnitude; for moderate positive $A_0$ there is a 
cancellation in the RGE evolution of $h_\Lambda$ \cite{Bernhardt:2008jz}.

As a first example, we show in Fig.~\ref{LSP_candidates}b 
the case of a right-handed $\tilde{\mu}_R$ LSP which we 
obtain via $\lambda_{132}|_{\rm GUT}=0.09$ \cite{Dreiner:2008ca}. 
We see that the $\tilde{\mu}_R$ LSP exists in an extended region of
${\not R_p}$ mSUGRA. We find a $\tilde{\mu}_R$ LSP for all $M_{1/2} 
> 480$ GeV, because $M_{1/2}$ increases the mass of the (bino-like)
$\tilde{\chi}_1^0$ faster than the mass of the $\tilde{\mu}_R$ 
\cite{Dreiner:2008ca}.

Our next example is a muon sneutrino $\tilde{\nu}_\mu$ 
LSP via $\lambda'_{231}|_{\rm GUT}=0.11$ \cite{Bernhardt:2008jz}, 
{\it cf.} Fig.~\ref{LSP_candidates}c. 
We again observe that the $\tilde{\nu}_\mu$ LSP exists in large regions of the 
mSUGRA parameter space. 
We can get a $\tilde{\tau}_1$ LSP instead of a 
$\tilde{\nu}_\mu$ LSP if we increase $M_{1/2}$, because the left-handed
$\tilde{\nu}_\mu$ LSP couples stronger to the gauginos than the (mainly right-handed) 
$\tilde{\tau}_1$. If we choose $M_0$ large enough we can always reobtain
the $\tilde{\chi}_1^0$ LSP. 

Finally, we show in Fig.~\ref{LSP_candidates}d that also squark LSPs are
possible via a non-vanishing baryon number violating operator, 
{\it cf.} Eq.~(\ref{superpot}). Fig.~\ref{LSP_candidates}d gives the example of 
a stop $\tilde{t}_1$ LSP via $\lambda''_{323}|_{\rm GUT}=0.35$ \cite{Dreiner:2008ca}.
We observe that the $\tilde{t}_1$ LSP parameter space is more restricted compared to
the aforementioned LSP candidates. We can also see in Fig.~\ref{LSP_candidates}d that
$A_0=\mathcal{O}(-1 {\rm TeV})$ is vital to obtain a $\tilde{t}_1$ LSP. 
On the one hand this $A_0$ increases left-right mixing. On the other hand it increases
the {\it negative} effect of the top Yukawa coupling and of $\lambda''_{323}$ 
on the running of the $\tilde{t}_1$ mass; see Ref.~\cite{Dreiner:2008ca}
for further details.

Beside the $\tilde{\tau}_1$, $\tilde{\mu}_R$, $\tilde{\nu}$ and $\tilde{t}_1$ we
also found the $\tilde{e}_R$, $\tilde{s}_R$, $\tilde{d}_R$ and $\tilde{b}_1$ 
as (the only possible) further non-$\tilde{\chi}_1^0$ LSP candidates in ${\not R_p}$ mSUGRA
\cite{Dreiner:2008ca}.

\section{Hadron Collider Phenomenology}

We have found in the last section many new candidates for the LSP in ${\not R_p}$
mSUGRA models. As an obvious next step, we will now investigate the hadron collider 
phenomenology of some of these new scenarios. 

The collider phenomenology of $\tilde{\tau}_1$ LSP scenarios is mainly driven by the
different decay modes of the $\tilde{\tau}_1$ LSP. It can either decay via a 
4-body decay or via a 2-body decay;
see Refs.~\cite{Allanach:2003eb,stau_pheno,Bernhardt:2008mz} for more details
and explicit examples. 

The $\tilde{\mu}_R$ LSP in Fig.~\ref{LSP_candidates}b will decay 
via $\lambda_{132}$, {\it i.e.} $\tilde{\mu}_R \rightarrow e \nu_\tau, \tau \nu_e$. 
One typical and simple supersymmetric process at the LHC will therefore be
\begin{equation}
PP \rightarrow \tilde{q}_R \tilde{q}_R \rightarrow 
(q\tilde{\chi}_1^0) (q\tilde{\chi}_1^0) \rightarrow 
(q \mu \tilde{\mu}_R) (q \mu \tilde{\mu}_R) \rightarrow 
(q \mu e \nu_\tau) (q \mu \tau \nu_e) \, .
\end{equation}

Even this simple supersymmetric process leads to four charged
leptons in the final state; two muons from the decay of the 
$\tilde{\chi}_1^0$ non-LSP into the $\tilde{\mu}_R$ LSP and
an electron and a tau from the $\tilde{\mu}_R$ LSP decays. 
Due to the large number of charged leptons, we
expect discovery of $\tilde{\mu}_R$
LSP scenarios at LHC to be relatively easy.

One of the most striking signatures of $\tilde{\nu}_\mu$ LSP scenarios, 
Fig.~\ref{LSP_candidates}c, are high-$p_T$ muons, {\it i.e.} muons with a
transverse momentum of a few hundred GeV \cite{Bernhardt:2008jz}.
At the same time we have a quark with roughly the same energy,
which is expected to be back-to-back to the muon. The high-$p_T$ 
muons stem from the decay of a (heavy) squark via $\lambda'_{231}$,
e.g. $\tilde{d}_R \rightarrow \mu t$. The large squark mass is in this case
transformed into the momenta of the standard model particles. 
High-$p_T$ muons can be found in roughly 10\% of all sparticle
pair production events at the LHC,
because $\lambda'_{231}=\mathcal{O}(g_i)$.

In general, we expect the discovery of squark LSP scenarios at the LHC to be more difficult
compared to ${R_p}$ conserving scenarios. Instead of large amounts of missing energy 
we have many jets in the final state from the LSP decays
\cite{Baer:1996wa}. However, it was claimed in Ref.~\cite{Choudhury:2005dg}
that the complete $\tilde{t}_1$ LSP region in Fig.~\ref{LSP_candidates}d 
should be testable at the Tevatron with the available data. But
this analysis has not been done so far.
 


\begin{thebibliography}{9}

  
\bibitem{Allanach:2003eb}
  B.~C.~Allanach, A.~Dedes and H.~K.~Dreiner,
  Phys.\ Rev.\  D {\bf 69} (2004) 115002.
  
\bibitem{Dreiner:2008ca}
  H.~K.~Dreiner and S.~Grab,
  Phys.\ Lett.\  B {\bf 679} (2009) 45.
  
\bibitem{softsusy}
  B.~C.~Allanach and M.~A.~Bernhardt,
  arXiv:0903.1805 [hep-ph].
  
\bibitem{Chun:1999cq}
  E.~J.~Chun and H.~B.~Kim,
  Phys.\ Rev.\  D {\bf 60}, 095006 (1999);
  H.~S.~Lee,
  Phys.\ Lett.\  B {\bf 663} (2008) 255.

\bibitem{Bernhardt:2008jz}
  M.~A.~Bernhardt, S.~P.~Das, H.~K.~Dreiner and S.~Grab,
  Phys.\ Rev.\  D {\bf 79} (2009) 035003.
  
\bibitem{stau_pheno}
  B.~C.~Allanach, M.~A.~Bernhardt, H.~K.~Dreiner, C.~H.~Kom and P.~Richardson,
  Phys.\ Rev.\  D {\bf 75} (2007) 035002;
    H.~K.~Dreiner, S.~Grab and M.~K.~Trenkel,
  Phys.\ Rev.\  D {\bf 79} (2009) 016002;
   B.~C.~Allanach, M.~A.~Bernhardt, H.~K.~Dreiner, S.~Grab, C.~H.~Kom and P.~Richardson,
  arXiv:0710.2034 [hep-ph].

\bibitem{Bernhardt:2008mz}
  M.~A.~Bernhardt, H.~K.~Dreiner, S.~Grab and P.~Richardson,
  Phys.\ Rev.\  D {\bf 78} (2008) 015016.

\bibitem{Baer:1996wa}
  H.~Baer, C.~h.~Chen and X.~Tata,
  Phys.\ Rev.\  D {\bf 55} (1997) 1466.
  
\bibitem{Choudhury:2005dg}
  D.~Choudhury, M.~Datta and M.~Maity,
  Phys.\ Rev.\  D {\bf 73} (2006) 055013.

\end{thebibliography}
\end{document}